\documentclass[aps,prd,twocolumn,nofootinbib,superscriptaddress]{revtex4}

\usepackage{color}
\usepackage{graphicx}
\newcommand{\be}{\begin{eqnarray}}
\newcommand{\ee}{\end{eqnarray}}

\def\nue{{\nu_e}}
\def\anue{{\bar\nu_e}}
\def\numu{{\nu_{\mu}}}
\def\anumu{{\bar\nu_{\mu}}}
\def\nutau{{\nu_{\tau}}}
\def\anutau{{\bar\nu_{\tau}}}

\newcommand{\sss}{\sin^2 \theta_{12}}
\newcommand{\sch}{\sin^2 \theta_{13}}

\def\ltap{\ \raisebox{-.4ex}{\rlap{$\sim$}} \raisebox{.4ex}{$<$}\ }
\def\gtap{\ \raisebox{-.4ex}{\rlap{$\sim$}} \raisebox{.4ex}{$>$}\ }

\begin{document}

\title{On the Observability of 
Collective Flavor Oscillations 
in Diffuse Supernova Neutrino Background}

\author{Sovan Chakraborty}
\affiliation{Saha Institute of Nuclear Physics,\ 1/AF Bidhannagar, Kolkata 700064,\ India}

\author{Sandhya Choubey}
\affiliation{Harish-Chandra Research Institute,\ Chhatnag Road, Jhunsi, Allahabad  211019\ India}

\author{Kamales Kar}
\affiliation{Saha Institute of Nuclear Physics,\ 1/AF Bidhannagar, Kolkata 700064,\ India}

\begin{abstract}
Collective flavor oscillations are known to bring multiple 
splits in the supernova (SN) neutrino and antineutrino spectra. 
These spectral splits depend not only on the mass hierarchy 
of the neutrinos but also on the initial relative flux composition. 
Observation of spectral splits in a future galactic supernova 
signal is expected to throw light on the mass hierarchy pattern of the 
neutrinos. However, since the Diffuse Supernova Neutrino 
Background (DSNB) comprises of a superposition of neutrino 
fluxes from all past supernovae, and since different supernovae are 
expected to have slightly different initial fluxes, it is 
pertinent to check if the hierarchy dependent signature of 
collective oscillations can survive this averaging of the flux spectra. 
Since the actual distribution of
SN with initial relative flux spectra 
of the neutrinos and antineutrinos is unknown, we assume a log-normal 
distribution for them.
We study the dependence of the hierarchy sensitivity to the mean and 
variance of the log-normal distribution function. We find that the 
hierarchy sensitivity depends crucially on the 
mean value of the relative initial luminosity. The effect of the width  
is to reduce the hierarchy sensitivity for all 
values of the mean initial relative luminosity. 
We find that in the very small mixing angle ($\theta_{13}$) limit considering only statistical errors even for very moderate values of variance, there is almost 
no detectable hierarchy sensitivity if the mean relative luminosities of $\nu_e$ and $\bar\nu_e$ are 
greater than 1.

\end{abstract}


\maketitle
\section{Introduction}                         \label{sec:introduction}

Neutrinos coming from supernova explosions can give rich information about 
the explosion mechanism as they are the only particles that come from regions
deep inside the core. They are also extremely useful for determining neutrino
properties as neutrinos from SN1987A have amply demonstrated
\cite{Bionta87, Alexeyev87, Hirata87}. 
In particular, it has been shown that the neutrino mass hierarchy and 
smallness of $\theta_{13}$ can be probed using 
SN neutrinos \cite{ds,lunardini-smirnov-0302033}. 
In the last few years, focus has been on the  
effect of neutrino-neutrino interaction 
in the central regions of the core of supernova, giving rise to 
the so-called collective effects \cite{Pantaleone:1992eq, Sigl:1992fn, Kostelecky:1994dt, 
Pastor:2001iu, Wong:2002fa, Balantekin:2006tg, Pastor:2002we, Sawyer:2005jk, 
Duan:2005cp, Duan:2006an, Hannestad:2006nj,Raffelt:2007yz, EstebanPretel:2007ec, Duan:2007mv, Raffelt:2007cb,
Raffelt:2007xt, Duan:2007fw, Duan:2007bt, Fogli:2007bk,
Fogli:2008pt,Duan:2007sh, Dasgupta:2008cd,  Dasgupta:2007ws,
Duan:2008za, Dasgupta:2008my, EstebanPretel:2008ni, Dasgupta:2008cu, Gava:2008rp, Raffelt:2008hr, Fogli:2008fj, Dasgupta:2009mg,
Fogli:2009rd, Chakraborty:09ej, Friedland:2010, Duan:2010bg, Dasgupta:2010a,
Dasgupta:2010b, Raffelt:2010b}. 

The `collective' nature of simultaneous flavor conversions of both neutrinos and
antineutrinos give rise to `splits' in the
spectra of the neutrinos and antineutrinos. These splits occur due to 
sudden change in the oscillation probability, causing spectral 
swaps which may end up in observable effects.  
Interestingly the impact of collective oscillations on the spectra
are different for the Normal Hierarchy (NH) and the Inverted
Hierarchy (IH). 
This opens up the possibility of identifying the 
neutrino mass hierarchy via observation of collective effects 
in the neutrino signal from a 
future galactic supernova event \cite{Dasgupta:2008my}. 

However with the very small rate of occurrence of galactic supernova events (a 
few per century) one is forced to think of strategies of detecting the 
above mentioned  effect otherwise. One of the promising possibilities is the 
detection of Diffuse Supernova Neutrino Background (DSNB) in the near future
\cite{Bisno:84,Krauss:1983zn,Woosley:86}. 
The cumulative number of neutrinos and antineutrinos produced by all earlier
SN events in the universe result in a cosmic background known as DSNB. 
Though these neutrinos are yet to be detected, 
one has reasons to believe that it may be 
possible to observe them in the near future \cite{Malaney:1996ar,Hartmann:1997qe,Totani:1995dw,Kaplinghat:1999xi,Strigari:2003ig,Fukugita:2002qw,Wurm:2007cy,Ando:2004hc,Cappellaro:1996cc,Ando:2002ky,Ando:2002zj,Lunardini:2005jf,Ando:2004sb,Volpe:2007qx,Cocco:2004ac,Strigari:2005hu,Beacom:2005it,Lunardini:2006sn,Chakraborty:2008zp,Galais:2009wi}. Present day upper limits
are 1.2 $\bar \nu_{e}$ cm$^{-2}$ s$^{-1}$ with energies above 19.3 MeV for
Super-Kamiokande (SK) water Cerenkov detectors \cite{Malek:2002ns}) and 
6.8 $\times 10^{3}$ cm$^{-2}$ s$^{-1}$ with energies between 25 MeV and 50 MeV
for Liquid Scintillator Detector (LSD) \cite{Aglietta:1992yk} at $90\%$ 
C.L. for 
both. 

The two key ingredients in the calculation of DSNB are (i) the SN rate which is
proportional to cosmic star formation rate and (ii) the $\nu$ and $\bar \nu$
energy spectra. Whereas reliable estimates are now available for the
star formation rate and the SN rate \cite{Hopkins-Beacom_06,Kistler_etal_07},
the prediction for the SN neutrino spectra has gone through an evolution
over the years. 
Earlier considerations of matter induced resonances were followed
by incorporating the `collective' effects due to 
interaction amongst the neutrinos themselves 
in the high density central regions of the core. 
The first study of the effect of the collective
flavor oscillations on DSNB fluxes and the corresponding 
predicted number of events in terrestrial detectors 
were carried out in \cite{Chakraborty:2008zp} where it was  
demonstrated that the event rate gets substantially 
modified by collective effects. The results also showed that 
observation of the DSNB fluxes at earth could shed light on 
the neutrino mass hierarchy. 
However, since then substantial progress has been made 
in the understanding of the collective effects. In particular, 
it is now clear that the neutrino and antineutrino survival probability and hence 
different split patterns depend crucially 
on the relative luminosities of the initial neutrino fluxes 
produced inside the exploding star \cite{Dasgupta:2009mg, Fogli:2009rd, Chakraborty:09ej}.
Therefore, one can predict the final neutrino and antineutrino 
spectra from a given SN with reasonable accuracy 
only if one already has access to the initial flux conditions. 
This complication is further compounded for the DSNB, 
as the DSNB flux comes from a superposition of the fluxes from 
all past SNe. Since the initial flux conditions are expected to 
be sensitive to the properties of the progenitor star and since 
we have a whole distribution of stars which end up being a SN, 
it is a complicated business to accurately estimate the DSNB spectra 
after accounting for the collective effects, which are bound to 
happen in almost every SN.


In this work we incorporate the observation of different split patterns in the
spectra for the calculation of DSNB and do not take the relative (anti)neutrino fluxes to
have fixed values. The main focus of this work is to check the 
effect of a distribution of supernovae with initial flux on the 
measurement of the neutrino mass hierarchy via the observation of 
the DSNB signal. Since the distribution of the initial fluxes over 
all past SNe are not available to us, we parametrize this 
by a log-normal distribution. The log-normal distribution has two 
parameters which define the mean and width of the distribution. 
Since they are also unknown, we choose various plausible values for them.  
We calculate the DSNB event rate averaged over these distributions. 
We study how the hierarchy measurement is affected when one takes the 
distribution of initial relative fluxes into account and find situations where the hierarchy 
determination may be possible.

\section{The diffuse supernova neutrino background}    \label{sec:dsnb}

The differential number flux of DSNB is
\begin{eqnarray}
F^{'}_{\nu}(E_\nu)
=\frac{c}{H_0}\,{\int\limits_0^{z_{max}}}\,R_{SN}(z)\,
F_{\nu}(E)
\frac{dz}{\sqrt{\Omega_m(1+z)^3+\Omega_\Lambda}}
~,
\label{eq:flux}
\end{eqnarray}
where $E_\nu=(1+z)^{-1}\,E$ is the redshifted neutrino energy 
observed at earth while $E$ is the neutrino energy produced at 
the source, 
$F_{\nu}$ is the neutrino flux
for each core collapse SN, $R_{\rm SN}(z)$ the cosmic SN
rate at redshift $z$, and the Hubble
constant taken as 
$H_0=70~h_{70} \,\mbox{km}~\mbox{s}^{-1}~\mbox{Mpc}^{-1}$. For the standard
$\Lambda$-CDM cosmology, we have
matter and dark energy density
$\Omega_{\rm m}=0.27$ and $\Omega_\Lambda=0.73$,
respectively~\cite{Amsler:2008zzb}. As Eq. (\ref{eq:flux}) suggests 
the DSNB flux at earth depends on two factors: (i) the cosmic SN rate 
and (ii) the initial SN neutrino spectrum from each SN.

The cosmic SN rate is related to the star formation rate $ R_{SF}(z)$, through
a suitable choice of 
Initial Mass Function (IMF) as 
${R_{SN}(z)}$=$ 0.0132 \times R_{SF}(z) {M_{\odot}^{-1}}$\cite{Baldry-Glazebrook, Bhattacharjee:1st}.
The IMF takes into account that only stars with masses larger than $8 M_{\odot}$
result in supernova explosion.
For the cosmic star formation rate per co-moving volume we take 

\begin{eqnarray}
R_{\rm SF}(z)= 0.32~f_{SN}h_{70}\frac{e^{3.4z}}{e^{3.8z}+45}
\frac{\sqrt{\Omega_m(1+z)^3+\Omega_\Lambda}}{(1+z)^{3/2}}
~,
\end{eqnarray}
where $f_{SN}$ is normalization of the order of unity and $R_{\rm SF}(z)$ is in units of ${M_{\odot}}\mbox{yr}^{-1}\mbox{Mpc}^{-3}$ \cite{Madau98a, Ando:2004hc}.
The initial SN neutrino spectrum emitted from the neutrinosphere is
parametrized in the form~\cite{Keil:2002in}
\begin{eqnarray}
F^{0}_{\nu}(E)&=&(\frac{L_{\nu}^{0}}{\bar E})\times(\frac{(1+\alpha)^{1+\alpha}}
{\Gamma(1+\alpha)\bar E}
\left(\frac{E}{\bar E}\right)^{\alpha}e^{-(1+\alpha)E/\bar E})\,\nonumber\\
&=&\phi^{0}_{\nu}\times\psi(E),
\end{eqnarray}
where $\phi^{0}_{\nu}$ is the total initial flux estimated for the initial
luminosity $L_{\nu}^{0}$ and average energy($\bar E$). The spectral shape also
depends on the energy distribution $\psi(E)$, which is parametrized by the
pinching parameter $\alpha$. In this study we use $\bar E_{\nue}$= 12 MeV,
$\bar E_{\anue}$= 15 MeV, $\bar E_{\nu_{x}}$= $\bar E_{\nu_{y}}$=18 MeV with
$\alpha_{{\nu}_x}$=$\alpha_{\bar{\nu}_x}=\alpha_{{\nu}_y}$=$\alpha_{\bar{\nu}_y}=4$ and $\alpha_\nue$=$\alpha_\anue=3$.
Here $\nu_{x}$ is a linear combination of $\nu_{\mu},
\nu_{\tau}$ and $\nu_{y}$ is the combination orthogonal to $\nu_{x}$; in our
case $\nu_{x}$ and $\nu_{y}$ has same flux hence $F_{\nu_x}$=$F_{\nu_y}$.
The average energies of the different flux types will also vary from 
SN to SN. However for simplicity, in this work we choose to keep the average 
energies fixed. We assume that $3\times 10^{53}$ erg of energy is 
released in (anti)neutrinos by all SNe.

The emitted spectrum $F_{\nu}(E)$ is processed by collective flavor oscillation
and MSW oscillation effects over the huge drop of matter density inside SN. 
The collective oscillations are over within a few 100 km from neutrinosphere 
whereas the MSW oscillation takes place in the region $10^{4}-10^{5}$ km \cite{Tomas:2004gr} for the
solar and atmospheric mass squared differences. As the collective and MSW
oscillations are
widely separated in space, they can be considered independent of each other \cite{Fogli:2008fj}. 
Thus the flux reaching the MSW resonance region already has the effects of the 
collective oscillations. 
This assumption may not hold in SN models \cite{Duan:2006an}, where at late times the MSW resonance and the collective effects can be simultaneous as there the matter density falls substantially compared to the early time. In the time integrated DSNB flux it can give rise to some corrections. However we have ignored these effects as in this work we followed matter profile from studies \cite{Fogli:2008fj} with larger matter density \cite{Schirato:2002tg,Tomas:2004gr} finding the two oscillations regimes to be mutually exclusive even at late times.

It has been seen that collective oscillations can
give rise to different split patterns of the neutrino spectra depending on the
initial relative flux of $\nu_e$ and $\bar\nu_e$ with respect to  flavor
$\nu_x$ or $\nu_y$, so we define
$\phi^{r}_{\nu_{e}}$=$\frac{\phi^{0}_{\nu_{e}}}{\phi^{0}_{\nu_{x}}}$ and
$\phi^{r}_{\bar\nu_{e}}$=$\frac{\phi^{0}_{\bar\nu_{e}}}{\phi^{0}_{\nu_{x}}}$
as measures of the relative fluxes \cite{Chakraborty:09ej}. 
The electron antineutrino flux beyond the collective region can 
swap to $x$ flavor
above some energy (single split) or can swap in some energy interval (double
split) or even can remain unchanged (no split) depending on the initial
relative flux $\phi^{r}_{\nu_{e}}$ and $\phi^{r}_{\bar\nu_{e}}$
\cite{Dasgupta:2009mg, Fogli:2009rd,Chakraborty:09ej}. 
DSNB is affected differently with these different oscillation scenarios. To
incorporate the effect of collective oscillations we work in a effective two flavor scenario with
single angle approximation \footnote{Multi-angle effects can give rise to kinematical decoherence among angular modes and smear the spectral splits \cite{Raffelt:2007yz}, however for spherical symmetry the single angle approximation i,e neutrino-neutrino interactions averaged along a single trajectory comes out to be a fine approximation as the multi angle decoherence in such a case is weak against the collective features \cite{EstebanPretel:2007ec,Dasgupta:2009mg,Duan:2010bf}}.

Recent papers \cite{Friedland:2010, Dasgupta:2010b} have explored the effect of three flavors on the outcome of the split patterns in collective oscillations. The three flavor results differ a bit from the two flavors only for IH and that too in a small region of the initial flux parameter space ($\phi^{r}_{\nu_{e}}$,$\phi^{r}_{\bar\nu_{e}}$), where single
split (in 3 flavor) appears instead of the double splits (in 2 flavor). 
However the observed single split in NH for this
region remains unchanged in the 3 flavor treatment \cite{Dasgupta:2010b}. Thus in a small part \cite{Fogli:2009rd,Chakraborty:09ej} of the parameter space ($\phi^{r}_{\nu_{e}}$,$\phi^{r}_{\bar\nu_{e}}$), where 2 and 3 flavor results differ, both IH and NH have similar split pattern (single split) for the 3 flavor evolution.  In other parts of the initial flux parameter space the split patterns remain the same. Therefore while averaging over the whole parameter space the corrections in the 3 flavor treatment compared to the 2 flavor one coming from the small region would not be appreciable. The main conclusions of this paper -- as we later see -- therefore remain largely independent of these small corrections.
Moreover the three flavor effects are also found to be sensitive to the neutrino-neutrino interaction potential, the evolution of a three flavor system effectively behaves like a two flavor one with the reduction of the neutrino-neutrino interaction potential by only one order \cite{Dasgupta:2010b}. Hence for such a smaller neutrino-neutrino potential the two flavor treatment of the collective evolution is reasonable and for larger values the two flavor treatment can be again be considered as a reasonable approximation when averaging over the whole initial flux parameter space is taken into account. 

In this effective two flavor treatment the collective oscillations involve only two flavors ($\nu_{e},\nu_{x}$), while the other flavor ($\nu_{y}$) does not 
evolve. Thus the Flux ($F^{c}_{\nu_{\alpha}}$) after the collective oscillations are given by

$
F^{c}_{\nue}= P_{c}F^{0}_{\nue} + (1-P_{c})F^{0}_{\nu_x}
$ , $
F^{c}_{\anue}= \bar P_{c}F^{0}_{\anue} + (1-\bar P_{c})F^{0}_{\bar\nu_x}
$. 
\vskip 0.2 cm

Similarly the quantities $F^{c}_{\nu_{x}}$ and $F^{c}_{\bar\nu _x}$ can be estimated from the relations

$F^{c}_{\nu _x } + F^{c}_{\nue} = F^{0}_{\nue} + F^{0}_{\nu_x}$ and $F^{c}_{\bar\nu _x} + F^{c}_{\anue} = F^{0}_{\anue} + F^{0}_{\bar\nu_x}$. 
\vskip 0.2 cm As for the other flavor $\nu_{y}$($\bar\nu_{y}$) there is no collective evolution, hence $F^{c}_{\nu _y (\bar\nu _y)}= F^{0}_{\nu_y(\bar\nu_y)}$ .
The quantities $P_{c}$ and $\bar P_{c}$ are the neutrino and antineutrino survival 
probability after the collective effect, respectively.
The only way $\nu_{y}$ can affect the final neutrino spectrum is by MSW transition. 
As we consider self-induced neutrino oscillations to happen independent of the MSW, so the pre-processed flux $F^{c}_{\nu_{\alpha}}$ and the unchanged `y' flavor will
undergo the traditional MSW conversions. 

In NH the MSW resonances affect the $\nue$ flux, while
the $\anue$ flux remains almost unaffected. However for IH, MSW affect the $\anue$ flux, and not the $\nue$ flux. Then these neutrinos get redshifted while traveling independently as mass-eigenstates until they reach earth, here they are detected as flavor eigenstates before or after having undergone regeneration inside the earth. The fluxes ($F_{\nu_{e}}$ and $F_{\bar \nu_{e}}$) arriving at earth after both the collective and MSW oscillation are given in Table I. 

\begin{table}[t]
\begin{tabular}{lcl}
\hline
Normal hierarchy \\
\hline
$F_{\nue}= s_{12}^2 P_{c}(\phi^{r}_{\nu_{e}},\phi^{r}_{\bar\nu_{e}},E) (2 P_{H}-1) (F^{0}_{\nue}-F^{0}_{\nu_x})$\\
~~~~~~~~~~~~~~~~~~~~~~$+ s_{12}^2 (1-P_{H}) (F^{0}_{\nue}-F^{0}_{\nu_x})+ F^{0}_{\nu_x}$.  \\

\vspace{0.1cm}
$F_{\anue}= c_{12}^2 \bar P_{c}(\phi^{r}_{\nu_{e}},\phi^{r}_{\bar\nu_{e}},E) (F^{0}_{\anue}-F^{0}_{\nu_x}) + F^{0}_{{\nu}_x}$. \\

\\


\hline
Inverted hierarchy \\
\hline
$F_{\nue}= s_{12}^2 P_{c}(\phi^{r}_{\nu_{e}},\phi^{r}_{\bar\nu_{e}},E) (F^{0}_{\nue}-F^{0}_{\nu_x}) + F^{0}_{{\nu}_x}$. \\
\vspace{-0.29cm}
\\

$F_{\anue}= c_{12}^2 \bar P_{c}(\phi^{r}_{\nu_{e}},\phi^{r}_{\bar\nu_{e}},E) (2 P_{H}-1) (F^{0}_{\anue}-F^{0}_{\nu_x})$\\
~~~~~~~~~~~~~~~~~~~~~~$+ c_{12}^2 (1-P_{H}) (F^{0}_{\anue}-F^{0}_{\nu_x})+ F^{0}_{\nu_x}$.  \\

\hline
\end{tabular}
\caption{Electron neutrino and antineutrino spectra
emerging from a SN.}
\label{fluxtable}
\end{table}
In Table I, 
$P_{H}$ is the effective jump 
probability between the neutrino
mass eigenstates due to the 
atmospheric mass squared driven 
MSW resonance. It takes a value between $0$ and
$1$ depending on the value of the mixing angle $\theta_{13}$. For 
$\theta_{13}$ large ($ \sch \gtap 0.01$) $P_{H} \simeq 0$, 
while for small $\theta_{13}$ ($\sch \ltap 10^{-6}$) $P_{H} \simeq 1$ 
\cite{Bandyopadhyay:2003ts}. 
For the smaller limit of $\theta_{13}$ the effective jump 
probability has a non-trivial dependence on energy and time, due to multiple
resonances [14, 19] and turbulence [16, 17, 18]. 
Since these effects occur in the small time-window when the
shock wave is in the resonance region one can neglect this sub-leading
effect in the time integrated observables like DSNB events.
The quantities $s_{12}^2$ and $c_{12}^2$ stand for 
$\sss$ (taken to be $0.3$ for numerical studies) and 
$\cos^2\theta_{12}$, respectively.  The neutrino and antineutrino survival 
probability after the collective effect are calculated numerically taking $\Delta m^{2}=3 \times 
10^{-3}$ $eV^{2}$ and a small effective mixing angle of $10^{-5}$.
However it should be noted that the collective probabilities $P_{c}$ and $\bar P_{c}$ are almost independent of the specific non zero values 
of the mixing angle \cite{Hannestad:2006nj,Duan:2007bt,Dasgupta:2009mg}. Thus the effective mixing angle in the matter for both the limits $ \sch \gtap 0.01$ and $\sch \ltap 10^{-6}$ will give rise to similar 
collective features. We checked explicitly that the results remain the same
when one varies the effective mixing angle by orders of magnitude. Hence for the collective evolutions we consider a representative value of $10^{-5}$ as the effective mixing angle for both the limits.
For further details of analysis of the 
collective effects we refer the reader to \cite{Chakraborty:09ej}. 
\footnote{Note that $P_{c}= \bar P_{c} = 1$ 
gives back the SN flux without collective 
oscillation \cite{ds}} 
As discussed before, $P_c$ and/or $\bar P_c$, which is a 
function of the neutrino energy $E$, show pattern of 
sudden change between 0 and 1, leading to sudden change in 
the neutrino and/or antineutrino spectra. Depending on the number 
of times the value of $P_c$ (and/or $\bar P_c$) changes, we can have 
more than one sudden swap of the neutrino (and/or antineutrino) 
spectra, referred to in the literature as multiple split 
\cite{Dasgupta:2009mg, Fogli:2009rd,Chakraborty:09ej}.
The split patterns are now known to be crucially dependent on the 
initial relative flux densities of the neutrino and antineutrino. 
The initial relative neutrino and antineutrino 
flux densities are expected to vary among different SN. 

\begin{figure}
\includegraphics[width=.8\columnwidth,angle=270]{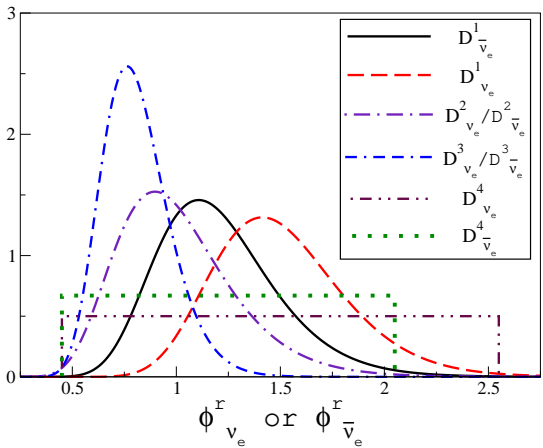}
\caption{\label{fig:Dist1-4}
 \footnotesize{
The four specimen distributions used in the calculations. The y-axis gives the distribution of supernova in arbitrary units.   
 }
}
\end{figure}

This variation in $\phi^{r}_{\nu_{e}}$ and $\phi^{r}_{\bar\nu_{e}}$ in
different SNe might have origin in difference of progenitor mass,
different luminosity etc. 
In fact even different simulations allow wide variation of
$\phi^{r}_{\nu_{e}}$ and $\phi^{r}_{\bar\nu_{e}}$ within the 2 fold
uncertainty around equipartition ($ \frac{ \bar E_{\nu_{x}}}{2 \bar E_{\nu_{e}}}\leq \phi^{r}_{\nu_{e}}\leq \frac{2 \bar E_{\nu_{x}}}{\bar E_{\nu_{e}}};\frac{\bar E_{\nu_{x}}}{2 \bar E_{\bar\nu_{e}}}\leq \phi^{r}_{\bar\nu_{e}} \leq \frac{2 \bar E_{\nu_{x}} }{\bar E_{\bar\nu_{e}}}~.$) \cite{lunardini-smirnov-0302033}. Most models  
predict $l_{\nu_{e}}\simeq l_{\bar\nu_{e}}$, where $l_{\nu_{e}}$ and $l_{\bar\nu_e}$ are the relative
luminosity ($\frac{L_{\nu_\alpha}}{L_{Total}}$) of $\nue$ and $\anue$ respectively \cite{Fogli:2009rd}. However, 
the combined luminosity of $\numu$, $\nutau$, $\anumu$ and $\anutau$  
is seen to be rather disparate between the different model 
results\footnote{See for e.g. the compilation of 
model results in \cite{Keil:2003sw}.}. 
The most reliable way to reconstruct the 
relative luminosity distribution 
function in principle would be from direct observation of 
SN events along with their neutrino signal. However, 
as yet only SN1987A has been observed along with the 
detection of its neutrinos/antineutrinos. We require to know neutrino 
fluxes of different flavors from a number 
of galactic SN with a range of
stellar mass and initial conditions before 
collapse to have information about 
the possible variation in $\phi^{r}_{\nu_{e}}$ and $\phi^{r}_{\bar\nu_{e}}$. 
This  might take decades and 
is clearly not possible in the near future. So we propose in this paper that 
the SN events contributing to DSNB have various different values of 
$\phi^{r}_{\nu_{e}}$ and $\phi^{r}_{\bar\nu_{e}}$. 
For quantitative estimates 
we assume specific distributions
for them. We take $\phi^{r}_{\nu_{e}}$ and $\phi^{r}_{\bar\nu_{e}}$  
distributed log-normally, defined by the parameters 
$(\mu_{1}, \sigma_{1}) $ and $(\mu_{2}, \sigma_{2}) $ i.e,
\be
D_{\nu}(\mu_{1}, \sigma_{1}) = \frac{e^{-(\frac{log(\phi^{r}_{\nu_{e}}) -
      \mu_{1}}{2 \pi \sigma_{1}})^{2}}}{\sqrt{2\pi}\sigma_{1}
  \phi^{r}_{\nu_{e}}}; \nonumber\\
D_{\bar\nu}(\mu_{2}, \sigma_{2}) =
\frac{e^{-(\frac{log(\phi^{r}_{\bar\nu_{e}}) - \mu_{2}}{2 \pi
      \sigma_{2}})^{2}}}{\sqrt{2\pi}\sigma_{2} \phi^{r}_{\bar\nu_{e}}}
~.
\label{eq:distr}
\ee

We choose a range of values for 
$\mu$ such that the expectation (mean) values of
$\phi^{r}_{\nu_{e}}$ and $\phi^{r}_{\bar\nu_{e}}$ are 
compatible with either Lawrence Livermore (equipartition) or 
Garching simulations. The parameter $\sigma$ determines the width 
of the distribution. Since the 
variation of $\phi^{r}_{\nu_{e}}$ and $\phi^{r}_{\bar\nu_{e}}$ is expected to
be within the 2 fold uncertainty around 
equipartition \cite{lunardini-smirnov-0302033}, the $\sigma
$'s for the distributions are chosen 
such that the distribution is well within the 2
fold uncertainty of the expectation value. In order to 
show the effect of the choice of the distribution function we 
simulate our results for four widely different distributions
for $\phi^{r}_{\nu_{e}}$ and
$\phi^{r}_{\bar\nu_{e}}$. They are chosen as follows:
\begin{enumerate}
 \item 
The expectation is the equipartition value, 
i.e 1.5 for $\phi^{r}_{\nu_{e}}$
and 1.2 for  $\phi^{r}_{\bar\nu_{e}}$ and the variation is within 2 fold
uncertainty of 1.5 and 1.2 respectively. The corresponding values of 
$(\mu,\sigma)$ turn out to be (0.39,0.21) for neutrinos and 
(0.16,0.24) for antineutrinos. 
We denote this distributions by $D^1_\nue$ and $D^1_\anue$.

\item
The expectation is same for both $\phi^{r}_{\nu_{e}}$ and
$\phi^{r}_{\bar\nu_{e}}$  and is taken as 1, the $\sigma$ is chosen as stated
above. Hence the distributions for $\nue$ and $\anue$ 
are identical and is defined by the parameters 
$(\mu,\sigma)\equiv (-0.03,0.28)$. 
We denote this distribution by $D^2_\nue$ and $D^2_\anue$.

\item 
The expectation is taken to be the same as the Garching simulations 
\cite{Keil:2002in}, i,e
0.8 for both the $\nue$ and $\anue$ and the distributions 
parametrized by $(\mu,\sigma)\equiv (-0.23,0.20)$ are the same.
We denote this distribution by $D^3_\nue$ and $D^3_\anue$.

\item
We consider the distribution to be constant between 0.45-2.55 
for $\phi^{r}_{\nu_{e}}$ and 0.45-2.05 for 
$\phi^{r}_{\bar\nu_{e}}$, with a normalization factor of 
$\frac{1}{2.1}$ and $\frac{1}{1.6}$ respectively. 
We denote this case by $D^4_\nue$ and $D^4_\anue$. 
\end{enumerate}
The four specimen 
distributions for $\phi^{r}_{\nu_{e}}$ and $\phi^{r}_{\bar\nu_{e}}$
are shown in Fig. \ref{fig:Dist1-4}. By definition the log-normal 
distribution functions $D^1$, $D^2$ and $D^3$ are normalized to unity. 
The choice of our normalization factors for the uniform distribution 
functions in $D^4$ ensures that they are 
normalized to unity as well. We can see from 
the figure that $D^3$ has the narrowest spread in the relative 
flux while the uniform distribution $D^4$ has the widest. For simplicity 
we have chosen distribution functions that are independent of the 
redshift $z$.

\begin{figure}[b]
\vskip -0.5 cm
\includegraphics[width=.75\columnwidth,angle=270]{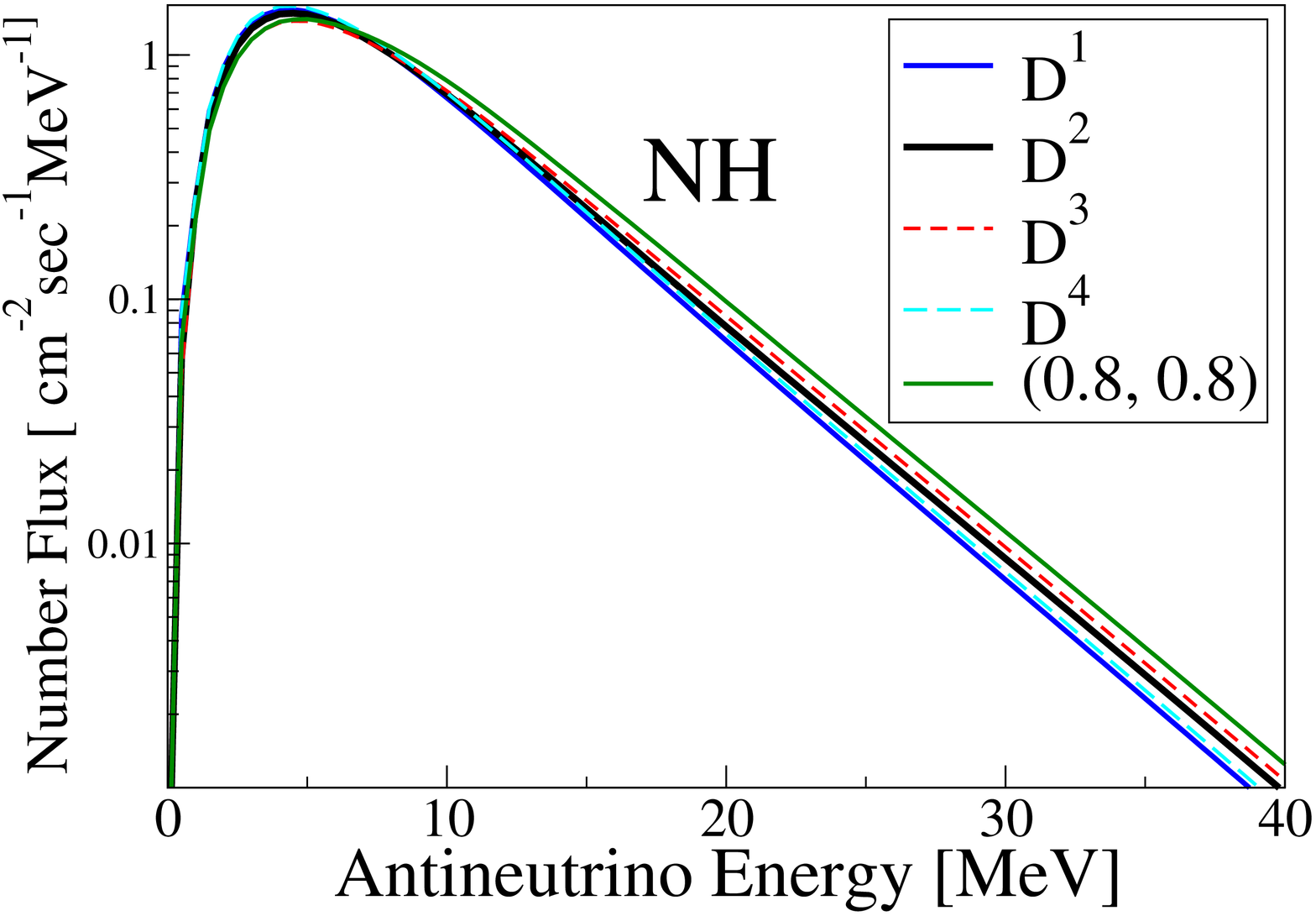}
\vglue -0.30cm \hglue 0.0 cm
\includegraphics[width=.75\columnwidth,angle=270]{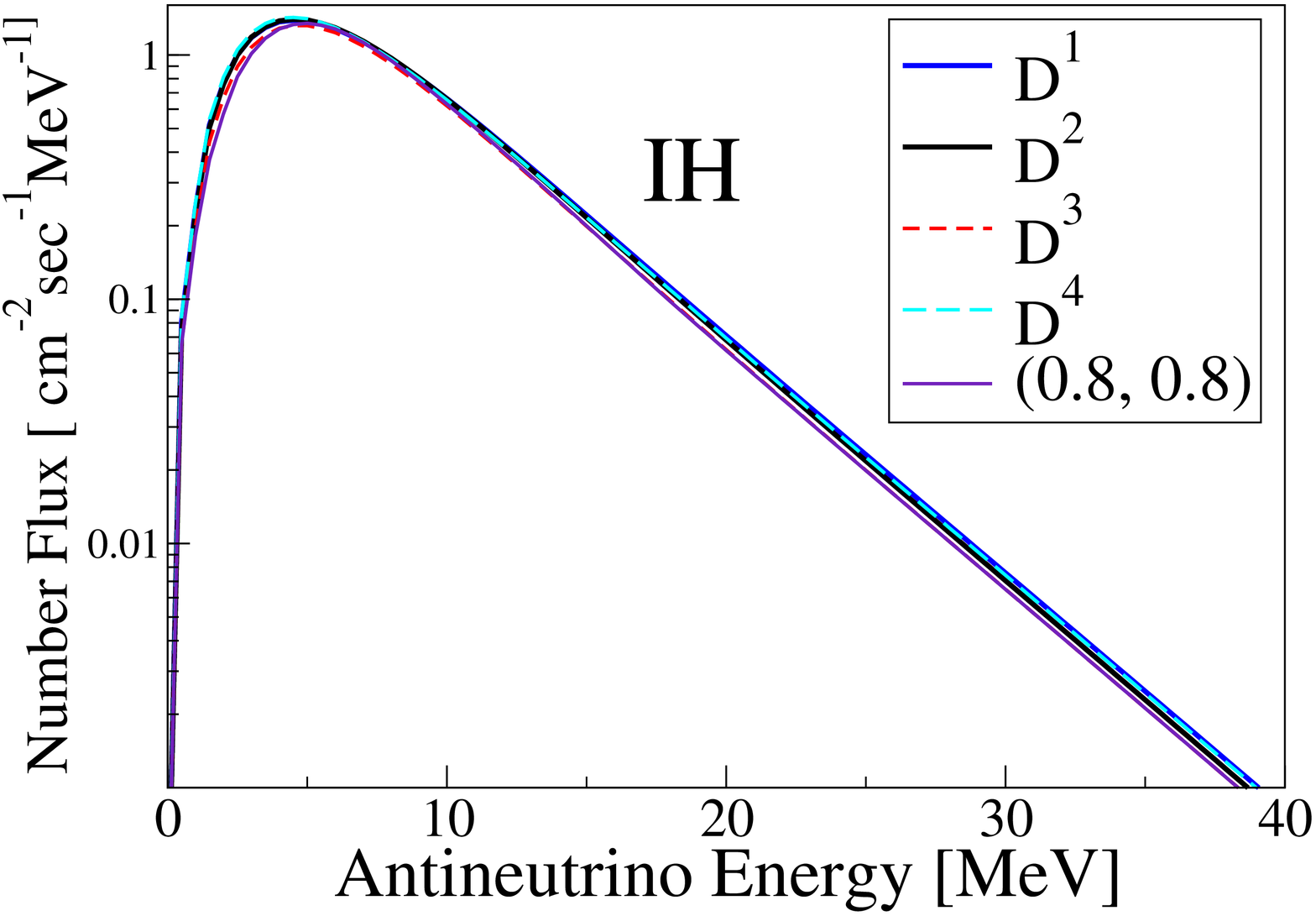}
\caption{$\bar\nu_e$ fluxes for NH(upper
panel) and IH (lower panel) with small $\theta_{13}$, i.e $P_{H}$= $1$.
\label{fig:dsn}}
\end{figure}

The differential number flux of DSNB 
with the $i^{th}$ distributions is given by
\begin{eqnarray}
F^{'i}_{\nu}(E_{\nu})=\frac{c}{H_0}\,{\int\limits_{\phi^{r}_{\nu_{e}min}}^
{\phi^{r}_{\nu_{e}max}}}\,{\int\limits_{\phi^{r}_{\bar\nu_{e}min}}^
{\phi^{r}_{\bar\nu_{e}max}}}\,{\int\limits_0^{z_{max}}}\,R_{SN}(z)\,
F_{\nu}(E,\phi^{r}_{\nu_{e}},\phi^{r}_{\bar\nu_{e}})\,\nonumber\\
\times
D^{i}_{\nu_{e}}(\phi^{r}_{\nu_{e}})\,\,D^{i}_{\bar\nu_{e}}
(\phi^{r}_{\bar\nu_{e}})\frac{dz~d\phi^{r}_{\nu_{e}}
  d\phi^{r}_{\bar\nu_{e}}}{\sqrt{\Omega_m(1+z)^3+\Omega_\Lambda}}
~,
\end{eqnarray}
where $ D^{i}_{\nu_{e}}(\phi^{r}_{\nu_{e}}) $ and $
D^{i}_{\bar\nu_{e}}(\phi^{r}_{\bar\nu_{e}}) $ is the number of supernovae in
between the interval
$\phi^{r}_{\nu_{e}}$ to $\phi^{r}_{\nu_{e}} + d\phi^{r}_{\nu_{e}}$ and
$\phi^{r}_{\bar\nu_{e}} $ to $ \phi^{r}_{\bar\nu_{e}} +
d\phi^{r}_{\bar\nu_{e}} $,
respectively. $F_{\nu}(E,\phi^{r}_{\nu_{e}},\phi^{r}_{\bar\nu_{e}})$ is the
neutrino flux of each SN with initial relative flux combination
$\phi^{r}_{\nu_{e}}$ and $ \phi^{r}_{\bar\nu_{e}}$ at a redshift z. $z_{max}$
is taken as 5, where as $\phi^{r}_{\alpha min}$ and $ \phi^{r}_{\alpha max}$
are chosen from the 2 fold uncertainty around expectation of the respective
distribution.
The upper panel in Fig.~\ref{fig:dsn} shows the antineutrino DSNB flux (in
logarithmic scale) for NH 
and the lower panel shows the flux for IH with $P_{H}$= $1$. In both panels
flux is shown for all of our different distribution ($D^1$-$D^4$) of
$\phi^{r}_{\nu_{e}}$ and 
$\phi^{r}_{\bar\nu_{e}}$. In Fig.~\ref{fig:dsn}  we
also plotted flux for the case without any distribution of
$\phi^{r}_{\nu_{e}}$ and $\phi^{r}_{\bar\nu_{e}}$ but specific value (0.8,
0.8) for the initial relative fluxes 
($\phi^{r}_{\nu_{e}}$,$\phi^{r}_{\bar\nu_{e}}$). 
From the flux figures it is evident that
the different distributions give similar DSNB flux. 
To check whether the small differences in their profile are measurable or not
we next calculate the corresponding event rate in different detectors.


\section{DSNB Event Rates and Hierarchy Measurement }
\label{sec:events}
\begin{table}
\begin{tabular}{|c|c|c|c|}
\hline
& &2.5 Mton-yr&GD+2.5 Mton-yr\\
Model&Hierarchy&(\footnotesize{19.3 - 30.0})
&(\footnotesize{10.0 - 30.0})\\
& & (MeV) & (MeV)\\
\hline
\hline
$D^{1}$&NH&76$\pm$9&280$\pm$17\\
\cline{2-4}
&IH ($P_{H}=0$)&92$\pm$10&319$\pm$18\\
\cline{2-4}
&IH ($P_{H}=1$)&80$\pm$9&288$\pm$17\\
\hline
$(1.5, 1.2)$&NH&75$\pm$9&281$\pm$17\\
\cline{2-4}
&IH ($P_{H}=0$)&98$\pm$10&332$\pm$19\\
\cline{2-4}
&IH ($P_{H}=1$)&75$\pm$9&280$\pm$17\\
\hline
\hline
$D^{2}$&NH&88$\pm$10&307$\pm$18\\
\cline{2-4}
&IH ($P_{H}=0$)&103$\pm$11&350$\pm$19\\
\cline{2-4}
&IH ($P_{H}=1$)&76$\pm$9&280$\pm$17\\
\hline
$(1.0, 1.0)$&NH&102$\pm$11&340$\pm$19\\
\cline{2-4}
&IH ($P_{H}=0$)&103$\pm$11&341$\pm$19\\
\cline{2-4}
&IH ($P_{H}=1$)&77$\pm$9&295$\pm$18\\
\hline
\hline
$D^{3}$&NH&98$\pm$10&334$\pm$19\\
\cline{2-4}
&IH ($P_{H}=0$)&113$\pm$11&379$\pm$20\\
\cline{2-4}
&IH ($P_{H}=1$)&70$\pm$9&258$\pm$16\\
\hline
$(0.8, 0.8)$&NH&112$\pm$11&378$\pm$20\\
\cline{2-4}
&IH ($P_{H}=0$)&113$\pm$11&378$\pm$20\\
\cline{2-4}
&IH ($P_{H}=1$)&70$\pm$9&260$\pm$17\\
\hline
\hline
$D^{4}$&NH&82$\pm$9&297$\pm$18\\
\cline{2-4}
&IH ($P_{H}=0$)&97$\pm$10&335$\pm$19\\
\cline{2-4}
&IH ($P_{H}=1$)&79$\pm$9&283$\pm$17\\
\hline
$(1.5, 1.25)$&NH&77$\pm$9&286$\pm$18\\
\cline{2-4}
&IH ($P_{H}=0$)&97$\pm$10&330$\pm$19\\
\cline{2-4}
&IH ($P_{H}=1$)&76$\pm$9&285$\pm$17\\
\hline
\end{tabular}
\caption{\label{tab:eventsa}
Number of expected events per 2.5 Megaton-year of a water Cherenkov with SK like resolution and in a similar detector with Gadolinium loaded.
}
\end{table}

Since the DSNB fluxes are very small, they are expected to be 
observed in either very large detectors or in reasonable size detectors 
with very large exposure times. The only reasonable size 
detector running currently 
is the Super-Kamiokande (SK) \cite{Malek:2002ns}. Amongst the proposed 
large detectors which have low energy threshold are the megaton water 
Cherenkov detectors \cite{hyperK, memphys, uno}, Gadolinium enriched water Cherenkov detectors \cite{Beacom:2003nk} 
(which could be SK or possibly an even larger water detector), 
very large liquid scintillator detector (LENA) 
\cite{Wurm:2007cy} or very large 
liquid Argon detector \cite{glacier, modular, flare}. 
Apart from the liquid Argon detector, none 
of the other detectors hold much promise for the detection of 
DSNB $\nue$. Therefore, in rest of this discussion we will focus 
only on the detection possibilities for the
DSNB $\anue$ fluxes. All detector technologies mentioned 
above use the inverse beta decay $ \bar\nu_{e} + p \longrightarrow n + e^{+} $
interaction for detecting DSNB fluxes. The only difference between 
them would be in terms of their energy window of sensitivity 
for the DSNB. 
Water detectors use the neutrino energy ($E_\nu$) window 19.3 MeV to 30.0 MeV 
\cite{Malek:2002ns},  whereas for  Gadolinium  
loaded water detectors the detection
window is between 10 MeV and 30 MeV 
\cite{Beacom:2003nk}. Liquid scintillator on 
the other hand has a sensitivity range between 
10 MeV and 25 MeV \cite{Wurm:2007cy}. 
We show the total number of events for 2.5 Mton-yr 
exposure in water detectors and 2.5 Mton-yr 
exposure in Gadolinium loaded water detectors 
in the third and fourth columns of Table II. 
The number of events are shown for the four different specimen 
distributions. 
Results are shown separately 
for $P_{H}=1$ and $P_{H}=0$ when the neutrino mass hierarchy is 
inverted (IH)
. For the NH, there is no 
dependence of the $\anue$ flux on $P_{H}$, as one can see from Table I.
In each case we also show for comparison the number 
of events expected when 
$\phi^{r}_{\nu_{e}}$  and $\phi^{r}_{\bar\nu_{e}}$ are {\it fixed} at their respective mean 
values of the distribution \footnote{For the uniform distribution $D^4$ we took the central values of the widths as their representative mean.}.
Observing a variation of the number of events changing the $\phi^{r}_{\nu_{e}}$ and $\phi^{r}_{\bar\nu_{e}}$ for these examples without any distribution is an important indicator of how inclusion of distribution can change the DSNB event numbers.

\begin{figure}
\includegraphics[width=.8\columnwidth,angle=270]{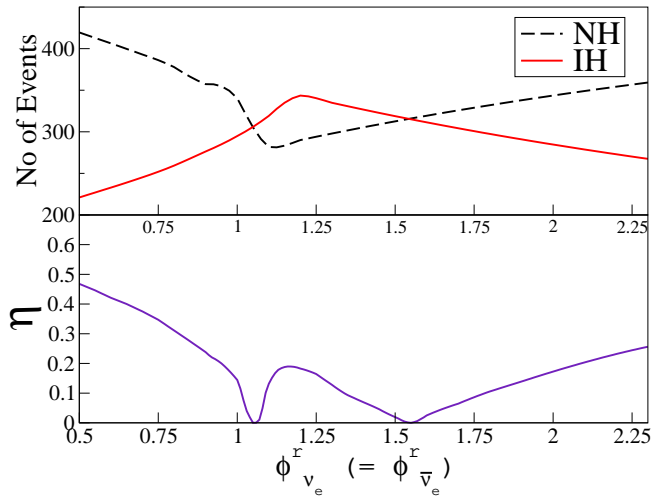}
\caption{\label{fig:ev}
 \footnotesize{
Upper panel shows the number of expected DSNB events in 2.5 MtonYr of Gadolinium loaded water 
detector for normal (NH) and inverted (IH) hierarchies (for IH, $P_{H} = 1$ ) as a function of 
  $\phi^{r}_{\nu_{e}} (= \phi^{r}_{\bar\nu_{e}})$. The lower panel shows the change of 
  $\eta$ as a function of $\phi^{r}_{\nu_{e}} (= \phi^{r}_{\bar\nu_{e}})$.
 }
}
\end{figure}

We next turn our attention to the possibility of measuring the 
neutrino mass hierarchy using DSNB detection. For the 
distribution examples $D^1$, $D^2$ and $D^4$ the difference of number of events between
NH and IH (both $P_{H}=1$ and $P_{H}=0$) seems to be within statistical uncertainty. Whereas
for the distribution $D^3$, though the difference in number of events 
between NH and IH (with $P_{H}=1$) decreases
significantly compared to the without distribution case, the
difference is still greater than the statistical uncertainty. 
To probe this variation of hierarchy sensitivity with the distribution further let us define the quantity
\be
\eta = \frac{\vert N_{NH} - N_{IH} \vert}{N_{NH} }
\,,
\ee
which gives a measure of the hierarchy sensitivity of the experiment. The quantities 
$N_{NH}$ and $N_{IH}$ are the number of expected DSNB events when the hierarchy is 
normal and inverted, respectively. The quantity $\eta$ is definitely not an observable as it depends on both
$N_{NH}$ and $N_{IH}$. However it helps describing the variation of hierarchy sensitivity of DSNB in a 
systematic manner.     
There are two ingredients of interest in Table II. 
Firstly, we can see that the relative hierarchy difference $\eta$ 
depends on the mean value of 
$\phi^{r}_{\nu_{e}}$  and $\phi^{r}_{\bar\nu_{e}}$. Secondly, for a given 
$<\phi^{r}_{\nu_{e}}>$  and $<\phi^{r}_{\bar\nu_{e}}>$ it also depends on the  
distribution function involved.

For a better understanding of the 
first issue, we show in the upper panel of Fig. \ref{fig:ev} 
the variation of the 
number of events for both the NH and IH cases, as a function of 
the relative luminosity factor. The analysis is again 
done for the small mixing angle limit ($P_{H}=1$) for IH, as it is the more 
challenging limit from the experimental point of view.  We show this for 
2.5 MegatonYears of Gadolinium doped water detector\footnote{Most of the results shown in this paper 
is for megaton class Gadolinium doped water detector to show the impact of the 
relative initial luminosity and its distribution on the hierarchy measurement using 
DSNB fluxes. The corresponding sensitivities for smaller scale detectors can be 
calculated trivially using these numbers.}.
For simplicity we have taken 
$\phi^{r}_{\nu_{e}} = \phi^{r}_{\bar\nu_{e}}$ in this figure and we do not take 
any distribution function into account. The lower 
panel shows the corresponding $\eta$ as a function of 
$\phi^{r}_{\nu_{e}} = \phi^{r}_{\bar\nu_{e}}$. We see that $\eta$ 
has a very complicated dependence on the relative initial luminosity functions. The variation of $\eta$ with $\phi^{r}_{\nu}$s actually depends on several factors like difference between split pattern of NH and IH, the split/swap energies, the initial relative flux of $\nu_e$, $\bar\nu_e$ and $\nu_x$.
It is rather high for very low values of $\phi^{r}_{\nu_{e}}$ and $\phi^{r}_{\bar\nu_{e}}$.  
It starts decreasing as the value of $\phi^{r}_{\nu_{e}}$ and $\phi^{r}_{\bar\nu_{e}}$ increase until 
it becomes zero around $\phi^{r}_{\nu_{e}} = \phi^{r}_{\bar\nu_{e}}\simeq 1.05$, thereafter 
it increases for a short while until it reaches a (local) maximum at around 
$\phi^{r}_{\nu_{e}} = \phi^{r}_{\bar\nu_{e}}\simeq 1.25$. 
Beyond that the value of $\eta$ decreases again reaching a second zero at 
around $\phi^{r}_{\nu_{e}} = \phi^{r}_{\bar\nu_{e}}\simeq 1.5$. After that it increases monotonically.

The most noteworthy thing in the upper panel of this 
figure is that the number of events for inverted hierarchy has almost linear dependence 
on $\phi^{r}_{\nu_{e}}=\phi^{r}_{\bar\nu_{e}}$ on both sides of the maximum which 
comes around $\phi^{r}_{\nu_{e}} = \phi^{r}_{\bar\nu_{e}}\simeq 1.25$. For the 
normal hierarchy one sees departure from linearity around $\phi^{r}_{\nu_{e}} = \phi^{r}_{\bar\nu_{e}}\simeq 1.1$. 
This feature is crucial in determining the effect of the distribution function on the 
hierarchy sensitivity. 
The effect of taking the distribution function into account boils down to 
creating a weighted average of the number of events, where the weights are determined by the 
distribution itself. For the log-normal case the weights are driven by the 
mean and width of the distribution, which we parametrize in terms of $\mu$ and $\sigma$. The 
effect of any distribution can thus be understood with the help of Fig. \ref{fig:ev}.

\begin{figure}
 \vskip 0.13 cm
\includegraphics[width=.75\columnwidth,angle=270]{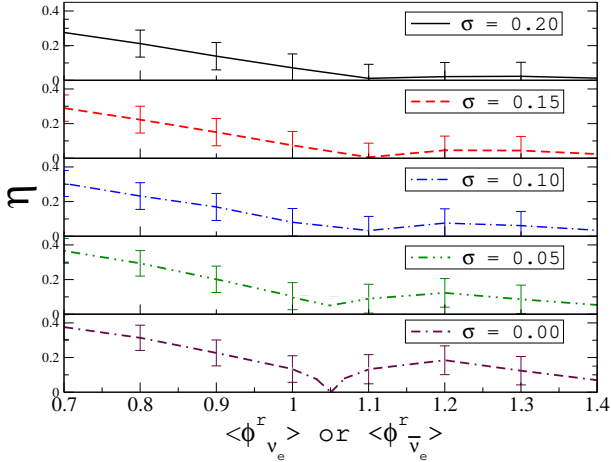}
\caption{\label{fig:error}
 \footnotesize{
Relative difference ($\eta$) of number of expected events in normal and inverted hierarchy, per 2.5 Megaton-year in a 
Gadolinium loaded SK like detector. For the IH case we consider small mixing angle i,e $P_{H} =$ 1. The above panels show $\eta$ for four possible cases with $\sigma$ being 0.00, 0.05, 0.10, 0.15 and 0.20. The x axis denotes the expectation of $\phi^{r}_{\nu_{e}}$ and $\phi^{r}_{\bar\nu_{e}}$ for the chosen sigma. Here expectation of $\phi^{r}_{\nu_{e}}$ and $\phi^{r}_{\bar\nu_{e}}$ are taken to be same i,e $\mu_1 = \mu_2 $. The errors shown are the statistical errors only.
 }
}
\end{figure}

To show the effect of the distribution function we continue to stick to the simplified scenario
where $\phi^{r}_{\nu_{e}} = \phi^{r}_{\bar\nu_{e}}$ and show in Fig. \ref{fig:error} the relative difference
$\eta$ as a function of the expectation value $\langle \phi^{r}_{\nu_{e}} \rangle(=
\langle\phi^{r}_{\bar\nu_{e}}\rangle)$. Here again we take  
$P_{H}=1$ for IH.
We consider log-normal distribution for the relative
luminosities 
{\footnote{Our conclusions remain fairly robust against the 
choice of the distribution function. We have explicitly checked this 
by repeating Fig. \ref{fig:error} with uniform distribution and normal distribution. However, 
we do not present those results as they follow the same pattern that we get for the 
log-normal distribution.}}
and show $\eta$ for five different values of $\sigma$ which controls the width of the
distribution. The values and error bars on $\eta$ correspond to a statistics of 2.5 Megaton-Year data
in Gadolinium loaded water detector.
The lowest panel with $\sigma=0$ corresponds to the case where we keep
$\phi^{r}_{\nu_{e}} ( \phi^{r}_{\bar\nu_{e}})$ fixed and for this case there is no effect of
the distribution. This case is similar to that in the lower panel of Fig. \ref{fig:ev}. We see that
without the effect of the distribution function almost all values of $\phi^{r}_{\nu_{e}} (= \phi^{r}_{\bar\nu_{e}})$
would give hierarchy sensitivity to at least 1$\sigma$ C.L., while for lower values of the relative luminosity the sensitivity can be seen to be rather good.
As we increase $\sigma$ the sensitivity is seen to go down for all value of the relative
luminosity. We find that even for very small values of $\sigma=0.05$, there is almost no
hierarchy sensitivity for $\langle \phi^{r}_{\nu_{e}} \rangle(=
\langle\phi^{r}_{\bar\nu_{e}}\rangle)\gtap 1$. As $\sigma$ increases this
pattern remains the same, though the sensitivity keeps falling for all values of
$\langle \phi^{r}_{\nu_{e}} \rangle(=
\langle\phi^{r}_{\bar\nu_{e}}\rangle)$.
In the above analysis, $\sigma$ is considered in a small range. The idea was to 
avoid deviating too much from the simulated values of the relative initial luminosities. If the actual variations of $\phi^{r}_{\nu}$s for all past supernovae are much larger than the range considered here, then 
the difference between predicted events for NH and IH would decrease even further, and this could washout the hierarchy sensitivity 
even for the low $\phi^{r}_{\nu}$ cases.
\begin{figure}
\vskip -0.65 cm
\includegraphics[width=.84\columnwidth,angle=270]{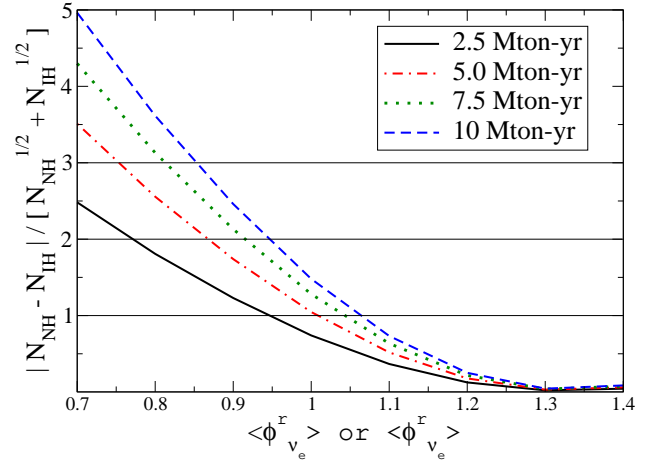}
\caption{\label{fig:difference}
 \footnotesize{ 
Plot of  $\frac{|N_{NH}-N_{IH}|}{\sqrt{N_{NH}}+\sqrt{N_{IH}}}$ with $<\phi^r_{\nu_e}>$ (= $<\phi^r_{\bar \nu_e}>$) for $\sigma$=0.20. The ratio is shown for a Gadolinium loaded SK like detector with different values of the exposure i.e. 2.5, 5.0, 7.5 and 10.0 Mton-Yr. Horizontal line at ratio 1 denotes that statistical error is of the same order as the difference between two hierarchies. Higher values like 2, 3 will be 
of more prominence as other uncertainties from flux, luminosities and detector systematics should be considered alongwith statistical error .
 }
}
\end{figure}

	Finally in Fig. \ref{fig:difference} 
     we display how we can make sure quantitatively that the
     observed number of events belong to one of the hierarchies and not the
     other one and see its dependence on the exposure in Mton-Yr. Fig. \ref{fig:difference}
     plots the ratio $\frac{|N_{NH}-N_{IH}|}{\sqrt{N_{NH}}+\sqrt{N_{IH}}}$ for
     four different values of the exposure i.e. 2.5, 5.0, 7.5 and 10.0 Mton-Yr
     as a function of the expectation values of the relative flux, 
     $<\phi^r_{\nu_e}>$ taken as equal to $<\phi^r_{\bar \nu_e}>$. This is
     for the case $P_{H}$=1 and with $\sigma$=0.20.
     This shows that with 2.5 Mton-Yr, 
     the relative flux expectations with values
     below 0.95 (ratio $\sim$ 1) can distinguish NH from IH (at least in principle) as the statistical error is smaller than the number of event difference.  However to achieve a better sensitivity i.e  ratio $\sim$ 2 the upper limit
     for the relative flux expectations is about 0.77. For ratio $\sim$ 3 the value is even lower.
     However with larger values of exposures like 5.0 or 7.5 Mton-Yr one can 
     still make a distinction between hierarchies when the centroid is below 
     0.75 and 0.82 respectively. But it is clear from the figure that by
     increasing the Mton-Yr one gets only a slow increase in the allowed
     parameter space.

\section{Conclusion}

In this work we studied the prospects of measuring the neutrino mass 
hierarchy from observation of the DSNB signal in terrestrial detectors. 
While such studies have been performed before by different groups 
including ours, our work is unique as this is the first time that 
distribution of the source SN 
with initial relative neutrino and antineutrino fluxes 
has been taken into account. It is natural that different 
SN would emit neutrino and antineutrino fluxes with slightly different 
initial conditions, depending on the properties of the progenitor 
star. This is particularly relevant in the context of collective 
oscillations, where the multiple split patterns depend crucially 
on the initial relative fluxes. Since the actual distribution function 
of SN with the initial relative fluxes are unknown, we chose four 
specimen distribution functions, which have a mean corresponding to 
the value from SN simulations and a width such that almost all the 
values are within a factor of two of the mean value. 
We worked with three log-normal and one uniform distribution. 
We presented the DSNB fluxes for all the four distributions for 
both normal and inverted hierarchies. We calculated the total predicted 
number of $\anue$ events in water detectors, both with and without 
Gadolinium. 
The log-normal distribution is characterized by its mean value and 
its variance. These are parametrized in terms of the variables $\mu$ and $\sigma$. 
We studied the dependence of the hierarchy sensitivity to the mean and 
variance of the log-normal distribution function. We concluded that the 
hierarchy sensitivity in this experiment had a crucial dependence on the 
mean value of the relative initial luminosity $\phi^{r}_{\nu_{e}}$ 
and $\phi^{r}_{\bar\nu_{e}}$. The sensitivity has a predominantly non-linear dependence on $\langle \phi^{r}_{\nu_{e}}\rangle $ and $\langle \phi^{r}_{\bar\nu_{e}}\rangle$, 
being 
higher for lower values of these quantities. The effect of the variance 
parametrized by $\sigma$ is to reduce the hierarchy sensitivity for all 
values of the mean $\langle \phi^{r}_{\nu_{e}}\rangle $ and $\langle \phi^{r}_{\bar\nu_{e}}\rangle$. 
We found that even for very moderate values of $\sigma \simeq 0.05$, there is almost 
no hierarchy sensitivity in the very small mixing angle limit for 
$\langle \phi^{r}_{\nu_{e}}\rangle = \langle \phi^{r}_{\bar\nu_{e}}\rangle \gtap 1$.
%

Finally we discuss the issue of using an effective two flavor evolution
followed in this paper. As mentioned earlier if the analysis is done using
all three flavors only the `double splits' for IH arising in a small part of 
the $(\phi^r_{\nu_e},\phi^r_{\bar \nu_e})$ parameter space seen in the two 
flavor case change to `single split' while things remain unchanged for NH. 
This effect would have been important for calculations for a single SN event
if the initial flux ratios fell in this part of the parameter space. However
for DSNB with averaging done over the SN distributions ($D_i$) the three flavor
treatment will result in small corrections to the effective two flavor results
presented here.

To summarize, using some reasonable simplifying assumptions we for the first
time take into account, for calculating DSNB, a distribution in the relative
fluxes in SN neutrinos and study its effect on the ability to distinguish the
two hierarchies through a future observation of DSNB. A more detailed and
rigorous study involving multi-angle effects and three flavors will be able to 
give more exhaustive answers to many of the aspects and should be persued in
future. But the model calculation carried out in this paper asserts that  
for DSNB which is a blend of many SN, not all identical, it is more difficult
to distinguish IH from NH compared to a single SN or the hypothetical case of
DSNB made of supernovae with identical flux ratios. A few favorable scenarios
for this hierarchy distinguishability are indicated. We feel that the central 
finding of this first study is robust and will survive more detailed analysis

\section*{Acknowledgments}
The authors would like to thank Basudeb Dasgupta for valuable comments and suggestions.
S. Chakraborty acknowledges hospitality at Harish-Chandra Research Institute 
during the development stage of this work.  S. 
Choubey acknowledges support from the 
Neutrino Project under the XI Plan of Harish-Chandra Research Institute. 
K. Kar and S. Chakraborty acknowledge support from the
projects `Center for Astroparticle Physics' and `Frontiers of
Theoretical Physics' of Saha Institute of Nuclear Physics.

\end{document}